\documentclass[conference]{IEEEtran}
\IEEEoverridecommandlockouts
\usepackage{cite}
\usepackage{amsmath,amssymb,amsfonts}
\usepackage{algorithmic}
\usepackage{graphicx}
\usepackage{textcomp}
\usepackage{xcolor}

\usepackage{enumerate}
\usepackage{subcaption}
\usepackage{siunitx}
\usepackage{algorithm}
\usepackage{stfloats}
\usepackage{caption}
\usepackage{multirow}
\usepackage{soul,color}
\usepackage{lipsum}
\usepackage{xurl}
\usepackage{nccmath}
\usepackage[symbol]{footmisc}
\usepackage{bm}
\usepackage{ulem}
\usepackage[final]{microtype}

\newcommand{\bcite}[1]{\hspace{-0.1pt}{\cite{#1}}}
\newcommand\redsout{\bgroup\markoverwith{\textcolor{red}{\rule[0.5ex]{2pt}{1pt}}}\ULon}

\def\BibTeX{{\rm B\kern-.05em{\sc i\kern-.025em b}\kern-.08em
    T\kern-.1667em\lower.7ex\hbox{E}\kern-.125emX}}
\begin{document}

\title{Towards the Avoidance of Counterfeit Memory: Identifying the DRAM Origin
}

\newcommand*{\affmark}[1][*]{\textsuperscript{#1}}
\author{\IEEEauthorblockN{
B. M. S. Bahar Talukder\affmark[$\ast$], Vineetha Menon\affmark[$\dagger$], Biswajit Ray\affmark[$\mathsection$], Tempestt Neal\affmark[$\ddagger$], and Md Tauhidur Rahman\affmark[$\mathparagraph$]}
\IEEEauthorblockA{
\affmark[$\ast\mathsection\mathparagraph$]\textit{Department of ECE, University of Alabama in Huntsville,} Huntsville, AL, USA \\
 \affmark[$\dagger$]\textit{Department of CS, University of Alabama in Huntsville,} Huntsville, AL, USA \\
 \affmark[$\ddagger$]\textit{Department of CSE, University of South Florida,} Tampa, FL, USA \\
 \{\affmark[$\ast$]bms.btalukder, 
 \affmark[$\dagger$]vineetha.menon, 
 \affmark[$\mathsection$]biswajit.ray, 
 \affmark[$\mathparagraph$]tauhidur.rahman\}@uah.edu,
 \affmark[$\ddagger$]tjneal@usf.edu}
 }


\maketitle

\begin{abstract}
 Due to the globalization in the semiconductor supply chain, counterfeit dynamic random-access memory (DRAM) chips/modules have been spreading worldwide at an alarming rate. Deploying counterfeit DRAM modules into an electronic system can have severe consequences on security and reliability domains because of their sub-standard quality, poor performance, and shorter life span. Besides, studies suggest that a counterfeit DRAM can be more vulnerable to sophisticated attacks. However, detecting counterfeit DRAMs is very challenging because of their nature and ability to pass the initial testing. In this paper, we propose a technique to identify the DRAM origin (i.e., the origin of the manufacturer and the specification of individual DRAM) to detect and prevent counterfeit DRAM modules. A silicon evaluation shows that the proposed method reliably identifies off-the-shelf DRAM modules from three major manufacturers.
\end{abstract}


\section{Introduction}\label{sec:introduction}
With the globalization of the semiconductor supply chain and the growth of the semiconductor market value, counterfeit integrated circuits (ICs) have become an established threat to the semiconductor community \cite{CounterfeitIC:UGuin, ChesDom, Counterfeit:CNET, Counterfeit:Micron, CSST_Rahman}. Counterfeit electronic parts and the risks associated with them have been increasing rapidly, which is reflected in the recent news \cite{CounterfeitIC:UGuin, ChesDom, Counterfeit:CNET, Counterfeit:Micron}. Many commercially available memory chips are fabricated worldwide in untrusted facilities and, therefore, a counterfeit memory chip/module can easily enter into the supply chain in different formats: recycled, re-marked, tampered, out-of-spec, forged-documented, defective, cloned, overproduced etc. \cite{CounterfeitIC:UGuin,ujjwalnew,ChesDom, foundaryIdentification:Wendt,ML:Ahmadi,forensic:Helinski}. Recent studies show that the global market share of counterfeit IC is worth $\$169$ billion, and $\sim 17 \%$ of which is contributed by memory chips \cite{ujjwalnew,ChesDom}.

The inclusion of counterfeit components in an electronic system can endanger personal and national privacy, sabotage critical infrastructure, and damage the viability of entire business sectors because of their sub-standard quality, poor performance, and a shorter life-span \cite{CounterfeitIC:UGuin,foundaryIdentification:Wendt,ML:Ahmadi,forensic:Helinski, ujjwalnew}. A counterfeit chip can fail any time after being deployed in the system, or they can be exploited to leak sensitive information or to allow remote access and endanger the integrity, confidentiality, and safety of a system by performing invasive or non-invasive fault-injection attacks \cite{rowhammer_PUF, rowhammer_DDR4}. Furthermore, recent experimental studies suggest that some DRAM modules are more vulnerable to rowhammer attack, a method of changing the restricted memory contents by repeated access to their adjacent rows \cite{rowhammer_DDR4}, because of their poor resiliency against noise, interference, etc.  

A single solution to detect or prevent counterfeiting is unrealistic because of the diversity of counterfeit types, sources, and refinement techniques \cite{CounterfeitIC:UGuin,foundaryIdentification:Wendt,ML:Ahmadi,forensic:Helinski}. Several discrete countermeasures have been proposed by the industry and academic researchers, which can be categorized into two major types: (i) electrical-based testing and (ii) physical-inspection based testing \cite{CounterfeitIC:UGuin,ChesDom, ujjwalnew, foundaryIdentification:Wendt,ML:Ahmadi,forensic:Helinski}. Some solutions are applicable to chips that are already in the market, and some solutions are integrated with the original chips for future tracking or metering \cite{CounterfeitIC:UGuin,ChesDom, ujjwalnew, foundaryIdentification:Wendt,ML:Ahmadi,forensic:Helinski}. Most of the solutions are ineffective for memory chips because they might require expensive equipment, expensive testing set-up, maintenance of an expensive database, and exhaustive enrollment process \cite{ChesDom,ujjwalnew, SCARE}. Besides, most physical-inspection based solutions are invasive and therefore, not applicable for mass-volume detection \cite{ChesDom}. For mass-volume verification and low-cost testing, electrical-based testing is required that is non-invasive \cite{ujjwalnew}. However, a non-invasive solution for counterfeit DRAM identification is extremely difficult because they may remain functional at the time of purchase and pass standard product qualification tests. Also, most existing solutions focus on a single counterfeit type (e.g., detecting recycled chips) \cite{CounterfeitIC:UGuin, ChesDom, SCARE}. Furthermore, current extensive regulations require a series of expensive testing methodologies \cite{ujjwalnew, ChesDom}. Therefore, a proper solution is required to identify counterfeit memory chips before deploying them in mission-, safety-, and security-critical systems.

In this paper, we propose a machine-learning-based technique to identify the origin of a DRAM manufacturer along with DRAM specification (i.e., density, grade, etc., see Sec. \ref{sec:ProbDef} for details) by exploiting DRAM latency (the required time to move charge from one location to another location in DRAM \cite{DRAMlatency:Chang,DRAMLatencyPUF:Kim, latencyTalukder,latencyTalukderTRNG}) variations to detect and prevent major counterfeit types. The major contributions of this paper include:
\begin{itemize}
\item We propose a framework to identify the origin of the DRAM manufacturer by exploiting the facts that the architectural, layout, and manufacturing process variations are reflected in latency variations. The framework is also capable of verifying specification of individual DRAM module.
\item We extract the most appropriate features from the latency-based erroneous patterns in DRAM modules to amplify the variations among manufacturers and specifications.
\item We propose a machine learning approach to determine the origin of the DRAM manufacturer based on the extracted features. The same method also separates DRAM modules of different specifications that are from the same manufacturer. 
\item We validate our proposed framework with off-the-shelf memory modules (commercial grade) from three major manufacturers- Micron, Samsung, and SK Hynix \cite{DRAMmarketShare}.
\item We validate the robustness of our proposed technique against temperature and voltage variations.
\end{itemize}
The rest of the paper is organized as follows. In Section \ref{sec:background-motivation}, we present the background of DRAM architecture, read/write operation, low-latency induced errors. We also present a set of motivations for our proposed framework in Section \ref{sec:background-motivation}. We highlight our major objectives and necessary assumptions in Section \ref{sec:ProbDef}. We propose the manufacturer identification framework in Section \ref{sec:method}. The experimental results are presented and discussed in Section \ref{sec:result}. We highlight the major limitations and the scopes of our proposed work in Section \ref{sec:futureWork}. We conclude our article in Section \ref{sec:conclusion}.

\section{Background and Motivations}\label{sec:background-motivation}
\subsection{DRAM Architecture and Latency Variations}\label{subsec:DRAMarchitecture}
A DRAM system can have one module or several modules depending on the memory requirement. A DRAM module is divided into one or multiple ranks \cite{DRAMlatency:Chang, DDR3:JEDEC, DRAM:Lui}. Each rank consists of several DRAM chips; and together, they provide a wide data-bus (usually 64 bits). The DRAM memory structure is analogous to a 2-D memory cell array. For simplicity, we can consider that each bit of the 64-bit word comes from 64 individual cell arrays. The rows of the DRAM are known as wordline (or page). The columns are known as bitline, and the chip density determines the total number of rows. The bitlines are connected to the row-buffer, a series of sense-amplifiers. Several DRAM cells are connected to a wordline. A DRAM cell consists of two components- an access transistor and a capacitor to hold the charge. The access transistor connects the capacitor with a bitline and is controlled by the wordline. The state of charge in the capacitor determines the memory content (i.e., `1' or `0'). Depending on the memory architecture, DRAM memory cells can be categorized into \textit{true-cell} and \textit{anti-cell} \cite{DRAM:Lui}. A \textit{true-cell} stores logic `1’ with a fully charged capacitor and `0' with an empty capacitor. On the other hand, an \textit{anti-cell} stores `0' with a fully charged capacitor and `1' with an empty state. However, the stored charge in the capacitor leaks away, which leads to an incorrect reading after a certain amount of time. So, to ensure the integrity, the content of a DRAM cell needs to be refreshed periodically before the memory contents flip. This time interval is known as the \textit{Retention} time, which is 32 or 64 milliseconds (ms) \cite{DDR3:JEDEC}. A failure to refresh before the retention time can alter the memory content.

To write/access the cell content, the column (or bitline) voltage need to be initiated at a specific reference voltage ($V_{ref} = \frac{V_{DD}}{2}$).
\begin{figure}[ht!]
\centering
\captionsetup{justification=centering, margin= 0.5cm}
\includegraphics[width=0.48\textwidth]{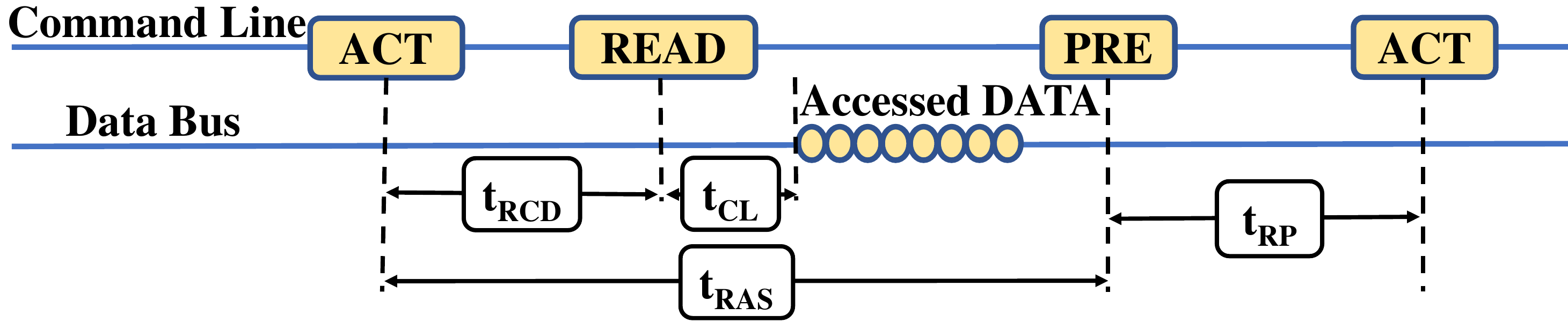} 
\caption{DRAM timing at reading cycle \cite{DRAMlatency:Chang}.}
\vspace{-3mm}
\label{fig:timing}
\end{figure}
Fig. \ref{fig:timing} presents a simplified version of DRAM read operation. A read operation starts with an ACTIVATE (ACT) command executed by the memory controller. $t_{RCD}$ is called the \textit{Activation} latency which is required to activate (turn `ON') the access transistor properly. The activated access transistor creates a conducting path between the storage capacitor and the bitline. The charge stored in the capacitor perturbs the bitline voltage. The sense-amplifier senses the perturbed voltage and amplifies it to an appropriate binary value. At this moment, the memory controller applies the READ command to read the data and fetch it to the data bus. The minimum time latency between the READ command and the first data bit to appear in data-bus is called \textit{Column Access Strobe} latency or \textit{CAS} latency ($t_{CL}$). DRAM's read operation is a destructive process. Therefore, the charge on the DRAM storage capacitor needs to be restored after each successful reading. The time required to activate an access transistor and to restore the charge on the corresponding storage capacitor is known as the \textit{Row Active} latency or \textit{Restoration} latency ($t_{RAS}$). At the end of the restoration process, the memory controller again applies the \textit{PRE} command to re-initiate all the bitlines for the next read/write operation. The \textit{PRE} command precharges all bitlines to $V_{ref}$. The time required to precharge all bitlines properly is called \textit{Precharge} Time ($t_{RP}$). The \textit{PRE} command also deactivates all previously activated access transistor. The $t_{RAS}+t_{RP}$ is the total time required to read a DRAM row properly; this total time is called \textit{Row Cycle} Time ($t_{RC}$). 

Usually, the DRAM manufacturer specifies a set of timing parameters for reliable read and write operation \cite{DDR3:JEDEC}. At the reduced timing latency below the standard value, we experience unreliable read and write operations, which is different from one module to another \cite{DDR3:JEDEC, DRAMlatency:Chang, DRAMLatencyPUF:Kim}. In our proposed method, we capture the architectural, layout and manufacturing process variations by exploiting the errors originated at the reduced \textit{Activation} latency ($t_{RCD}$).

\subsection{Counterfeit, Existing Work, and Motivations}\label{subsubsec:Motivations}
\begin{figure}[ht!]
\centering
\captionsetup{justification=centering, margin= 0.5cm}
\includegraphics[width=0.45\textwidth]{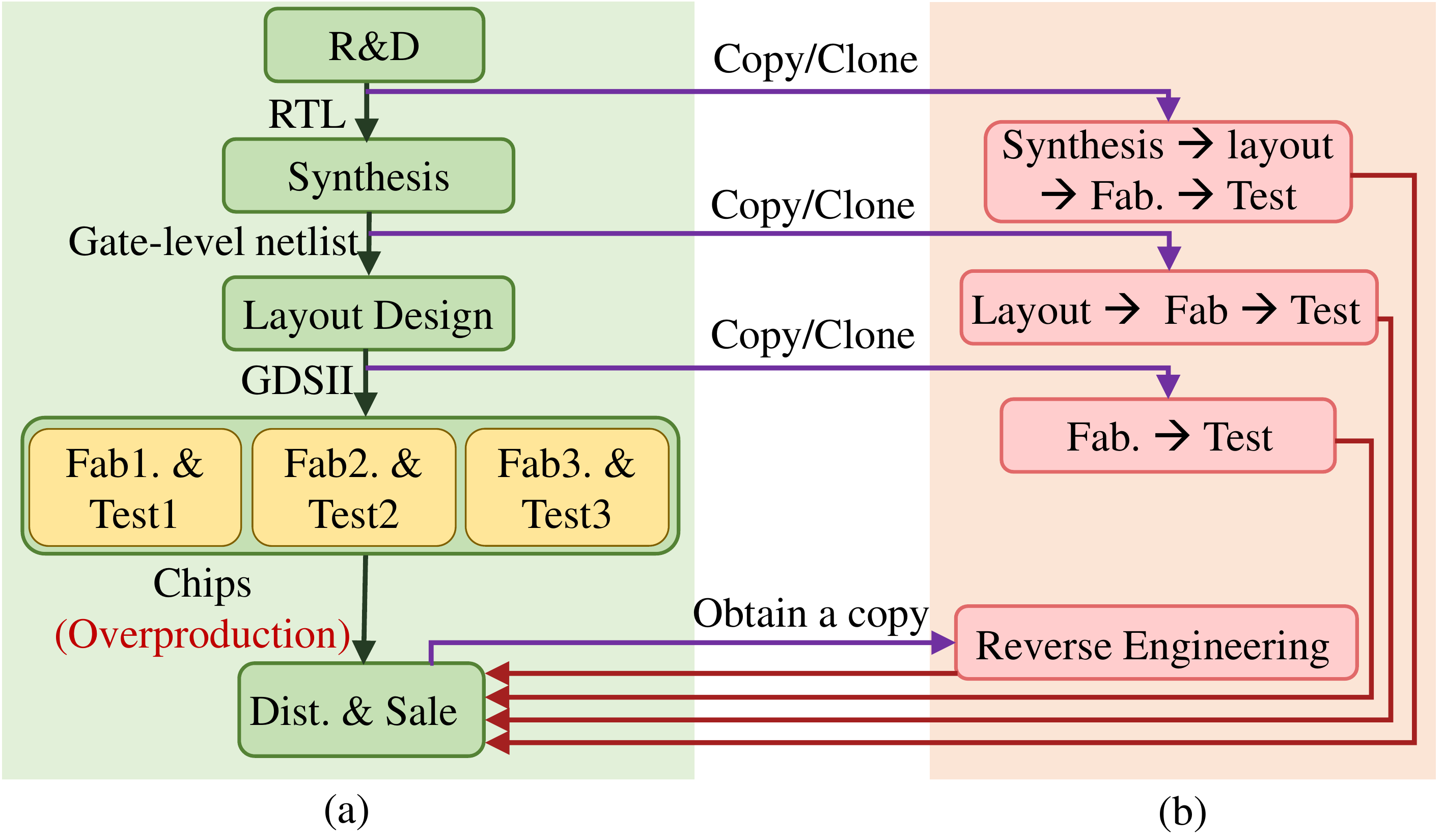} 
\caption{(a) Authentic-chip production cycle vs. (b) counterfeit-chip production cycle \cite{PVal:Basak}.}
\vspace{-3mm}
\label{fig:Piracy}
\end{figure}

The modern horizontal semiconductor supply chains involve several parties to reduce the fabrication cost and time to market \cite{CounterfeitIC:UGuin, ChesDom, ujjwalnew, supplyVLSI:Lui, PVal:Basak}. In this model, a chip is designed in one place while fabricated in a different place. Because of traveling IPs in different formats and involvement of untrusted parties, the modern semiconductor supply chain suffers from counterfeiting (such as hardware trojan or malicious change in third-party IP or chip layout, cloning IPs/ICs, remarking, etc.) \cite{CounterfeitIC:UGuin, PVal:Basak, CSST_Rahman,trojannima}. Fig. \ref{fig:Piracy} shows the IC/DRAM design flow for authentic-chip and pirated-chip production cycle. The untrusted party (the third-party IP developer, the foundry, the assembly, the distributor, etc.) can perform counterfeiting at different phases of the manufacturing process. An untrusted party can send out overproduced, and out-of-spec/defective ICs/DRAM chips to the market. The untrusted party also can clone the original chip by stealing IPs or by reverse-engineering (from a post-fabricated product) to avoid research and development (R\&D) costs. Recycled ICs also can be added to the supply chain at different stages (e.g., in the foundry or the assembly). Below, we summarize (but not limited to) the motivations of our proposed work. 

\hspace{13pt}\textbf{i) Existing countermeasures and their limitations:} In 2015, the Department of Defense imposed several rules to stop entering counterfeit components in the defense supply chain \cite{ujjwalnew}. These rules create more accountability and require proper testing standards along with maintaining a database if the electronic components are acquired from the untrusted party. Researchers, industries, and several organizations have developed several test plans or standards that require a series of testing methods. Visual inspection, X-ray imaging, scanning electron microscopy (SEM), energy disruptive spectroscopy, terahertz spectroscopy, etc., are the most common physical inspection-based techniques that are used to detect counterfeit chips \cite{CounterfeitIC:UGuin, foundaryIdentification:Wendt, ML:Ahmadi, forensic:Helinski}. Some imaging techniques capture the internal features of a sample and might be very effective to identify certain counterfeit types. However, these physical inspection-based techniques usually require expensive tool, long test time for sample collection and imaging, subject matter experts \cite{CounterfeitIC:UGuin}. Curve trace testing (CTT), parameter testing, burn-in testing, aging-based analysis, etc. are the most common electrical test methods that can be used to identify counterfeit components \cite{onChipSensor:Zhang, onChipSensor:He}. Existing burn-in and aging analysis-based techniques partially destructive because months to years of the device are consumed from the accelerated aging. The parameter testing is useful but requires expensive test setup and complex test programs. On the other hand, the CTT (electrical testing to capture damaged packages, broken or damaged wires, cracked die, etc.) is only useful for detecting recycled ICs \cite{ChesDom,ujjwalnew}.

Researchers also have proposed pre-fabrication techniques to prevent counterfeiting such as hardware metering (\bcite{CounterfeitIC:UGuin, hwmeter:Alkabani}), secure split test (SST) (\bcite{CSST_Rahman}), placing on-chip sensors (\bcite{onChipSensor:Zhang, onChipSensor:He, ChesDom, CDIR:Zhang}), electronic chip ID (\bcite{CounterfeitIC:UGuin}), DNA marking (\bcite{DNA:Hayward}), RFID-based tracking (\bcite{RFID}), blockchain-based traceability (\bcite{Kundu}), PUF-based techniques (\bcite{rowhammer_PUF, DPUF:Sutar, DRAMLatencyPUF:Kim, DPUF:Tehranipoor, Xiong:DPUF, Hashemian:DPUF}), etc. Hardware metering and SST are used to provide post-fabrication control but require hardware changes and complex supply chain management. Besides, each of the chips needs to go through the unlocking process, which is tedious and adds extra steps in the supply chain. On the other hand, on-chip sensors are used to monitor the device degradation but need some additional sensors and monitoring units. ECID tags each chip with a unique ID by adding some non-programmable memory (such as OTP or ROM). The ECID-based solution is susceptible to tampering. In DNA marking technique, a mixed plant DNA is used to create a new DNA sequence \cite{DNA:Hayward}. Later, this sequence is applied with the ink to mark the chip package. However, this technique suffers from a complex authentication scheme. For the PUF-based technique, the unique ID generated from each chip can be used to identify authentic chip. However, the PUF-based techniques require an extensive database for registration purposes. The registration and authentication procedures are exhaustive as well. The DRAM memory itself can be used as a PUF, but the uniqueness of PUF can make the solution very challenging. Moreover, most of the existing solutions address a single counterfeit type.

Our proposed machine-learning based technique has some advantages compared to other possible manufacturer identification techniques:
\begin{itemize}
    \item In the proposed machine-learning based technique, we store minimal information (a few statistical parameters) to attest and detect a large group of memory modules/chips. On the other hand, for example, in PUF-based technique, we need to create a golden/reference dataset by accumulating all challenge-response pairs for an individual memory module/chip.
    \item Our proposed technique does not need any exhaustive registration processes. Analyzing a small number of samples from a large group of modules is sufficient enough to reveal the detail statistical information of that group.
    \item Furthermore, the machine-learning based technique can serve some particular purpose; for example, the user might want to test the authenticity of the purchased device without knowing any details of chip design. In such a case, the user does not need to have exclusive access to any sensitive information (e.g., the challenge-response database of PUF-based scheme).
    \item This technique does not require any hardware modification unlike many existing techniques such as: hardware metering, SST, on-chip sensor-based authentication, ECID, etc. \cite{CounterfeitIC:UGuin}.
    \item The proposed solution is invasive and less expensive compared to some other techniques.
\end{itemize}

\textbf{ii) Importance of attesting the DRAM origin}: Attesting the origin of the foundry or the manufacturer is a critical need to (a) enforce the license agreement, (b) ensure and track the quality of the chip, (c) rank the manufacturer based on the quality and durability of their products, (d) protect the intellectual property, (e) ensure accountability, and (f) stop the spread of counterfeit memory chips worldwide. Counterfeiters usually hide the origin of the foundry, fake the quality of the original memory modules/chips, and infringe on the rights of the original manufacturer \cite{Counterfeit:CNET, Counterfeit:Micron, foundaryIdentification:Wendt, ML:Ahmadi, forensic:Helinski, CSST_Rahman}. On the other hand, verification of individual DRAM module's specification can detect certain counterfeit types such as upgrading a DRAM module through remarking or forged documentation.

The existing techniques on attesting of the foundry rely on simulation or test data from fabrication and packaging facilities \cite{foundaryIdentification:Wendt, ML:Ahmadi}. Unfortunately, in most cases, the testing data are not made publicly available and, therefore, a party that does not have access to those test data cannot make the classifier and (or) cannot verify the ratified foundry. In contrast, in our proposed technique, the DRAM chips can be authenticated based on trained classifier provided by the manufacturer or a trusted third party. In the proposed technique, the verifier does not require any prior knowledge of the manufacturing process. The classifier input, a set of features, can be easily evaluated in any low-cost embedded or FPGA-based system.

\textbf{iii) Vulnerability of counterfeit memory chips:} A counterfeit memory module can be more vulnerable to attacks because of their less resiliency against noise. For example, recent studies demonstrate that some DRAM modules possess inferior quality than others, which are more susceptible to rowhammer attack \cite{rowhammer_DDR4}. Our experimental results also suggest that a recycled DRAM chip is $\sim8\%$ more vulnerable to rowhammer attack.

\textbf{iv) Modification of serial presence detect (SPD) information:} DRAM manufacturers provide the information of all DRAM timing parameters in a small read-only memory (ROM) which is integrated with the DRAM module \cite{JEDEC:SPD}. It has been demonstrated that a counterfeiter can modify this information (aka SPD information) to make a DRAM authentic or superior \cite{SPD:zak}. These DRAM chips can pass the initial test but can cause critical failure of the system during runtime under some operating conditions that are considered safe \cite{reducedVoltage:chang}.

\section{Objectives and Assumptions} \label{sec:ProbDef}
The objective of this work is to identify the DRAM origin (i.e., the origin of the manufacturer and the specification of individual DRAM) reliably by capturing all variabilities. The major variations to attest and identify the origin of the manufacturer include:
\begin{itemize}
    \item \textbf{Architectural variations:} Manufacturers usually optimize the DRAM architecture in various ways to support the specifications and product cost \cite{DRAMArch:Jacob, LatVar:Lee}. For example, manufacturers shrink the die size to reduce the cost per cell, which can cause the DRAM more susceptible to noise, more vulnerable, and less robust. The minimum required latency parameters vary from one architecture to another. Therefore, for a given reduced timing latency, DRAMs from different manufacturers create a different amount of errors \cite{DRAMlatency:Chang,DRAMLatencyPUF:Kim, LatVar:Lee}.
    \item \textbf{Layout variations:} Chip layout variation from one manufacturer to another may originate from several sources such as chip area, floorplanning, placement, and routing, etc. \cite{LatVar:Lee, layoutVar:clein}. This layout variation may affect different electrical characteristics such as RC path delay, power utilization ($I^2R$ loss), noise margin, etc., which can be reflected in the DRAM latency parameters \cite{DRAMlatency:Chang,DRAMLatencyPUF:Kim}.
    \item{\textbf{Process variations:} The intrinsic process variation can be either random or systematic \cite{processVar:Cao, processVar:kuhn}. The random process variation can be considered as noise and varied among the chips fabricated on a single silicon wafer. On the other hand, the systematic process variation depends on the quality of the fabrication plant and also related to the microstructural locality and pattern. The process variation can lead to different error patterns at a given reduced latency parameter and can reveal the information about the fabrication plant and the microstructure.}
\end{itemize}

Our proposed technique of identifying the memory manufacturer is based on the following assumptions.
\begin{itemize}
    \item \textbf{Assumption on memory class}: A manufacturer ships memory module with a part number on it, which contains the manufacturer information and chip specification, such as density, speed grade, package, temperature range, bus width, die generation, etc. \cite{hynix:partnmbr}. In addition to part-number, a manufacturer also provides additional module specification on the module label \cite{hynix:Tchspprt}, such as JEDEC PCB layout version \cite{DesignFile:JEDEC}, SPD version \cite{JEDEC:SPD}, manufacturing country, manufacturing lot number, etc. Timing parameters of the DRAM module are specified into SPD data \cite{JEDEC:SPD}. In this work, two DRAM modules are considered as two different classes, if one of the following information is mismatched: i) manufacturer, ii) part number, and iii) PCB layout version/SPD data. A change in one specification can lead to different GDSIIs (related to the die generation, and specification), packaging, PCB layouts, or SPD data. Note that a manufacturer may send a single GDSII file to different fabrication plants. We assume that fabrication plants with the same GDSII follow similar design rules to minimize the effect of systematic process variations.

    \hspace{1.5em} In this article, `sample', `positive sample', and `negative sample' mean memory module under test, memory module that originally belong to the target class, and memory modules that originally do not belong to the classifier target class (i.e., belong to the outlier region), respectively. 

 \item \textbf{Assumption on data training, and verification}: We extract different features to capture the architectural, layout, and process variation. These features are trained to learn a statistical model. In our proposed scheme, the manufacturer or a trusted party is responsible for training the statistical model and releasing it for public use. We also assume that, while training the statistical model for a particular memory class, manufacturers do not have any statistical information from other memory classes (i.e., from negative class). However, in practice, the manufacturer may collect a few random samples from other classes. Training statistical model with some negative examples might be beneficial, but it is almost impossible to collect all samples from all negative classes because of the diversity of memory chips/modules and several manufacturers. We also assume that the chips/modules that are used for the enrollment in the proposed technique are authentic. The regular consumers should able to verify their purchased DRAM class by only using the statistical model. We also assume that consumers do not have any knowledge on memory architecture and manufacturing process.
\end{itemize}

\section{Proposed Method} \label{sec:method}
To extract all possible variabilities, we reduce the DRAM timing latency and obtain the signatures (i.e., the error pattern or fail bit count) that reflect the architectural, layout, and process variations. Below, we present a framework to identify the DRAM class that involves several steps. 

\textbf{Step 1: Data acquisition.} The experimental results show that the latency-induced error pattern depends on the data written into the memory, the amount of latency reduction, and the DRAM module \cite{PARBOR:Khan}. To capture the maximum variations among the memory classes, we write four different sets of data to the memory module: (i) Data set 1: solid data pattern (all 1's), (ii) Data set 2: inverse solid data pattern (all 0's), (iii) Data set 3: column stripe data pattern (101010$\cdots$), (iv) Data set 4: inverse column stripe data pattern (010101$\cdots$). Then, each data set is read back from the DRAM module at the reduced \textit{Activation} time ($t_{RCD}$) to capture module dependent erroneous outputs.

\textbf{Step 2: Feature selection.} It is crucial to select the optimum number of features since the performance of classifiers is sensitive to the choice of the features and features' attributes such as correlation, noise, and other factors. In this step, we will select the key features that can effectively capture architectural, layout, and process variations observed in DRAM since they directly impact the accuracy, computation time, and storage (of golden data) of our proposed technology. The classification models are created based on a total of 26 features collected from the four sets of data. Features are extracted from the whole data that is read out from one page. A single bank from 1GB memory module contains 16k+ pages per bank (eight banks per module), and for the case of a 2GB memory module, each bank includes 32k+ pages. Each memory page contains 1,024 words and each word contains 64-bits of data. The data collected (at reduced $t_{RCD}$) from each memory pages are then rearranged into a 1,024$\times$64 (denoted as $d_R$ of size $w\times b$, where, $w=$ number of words in a page, $b=$ number of bits in each word) binary array. Moreover, for each page, we create another array, $d_F$ (same size of $d_R$) which tracks the location of flipped bits. Note that, the $d_F(i,j)=1$, if $d_R(i,j)$ is flipped with respect to the actual data that is written to the DRAM otherwise, it will be 0. The following features have been chosen from each page to identify the DRAM origin (i.e., the origin of the manufacturer and the specification of individual DRAM).\\
\textbf{\textit{Feature 1 ($\Psi_1$)}:} The total number of flips, also known as failed bit count (FBC), is used to capture the data dependency, process variation, and layout variation of the DRAM chips. The silicon results show that the FBC counts change from one DRAM module to another module.\\
\textbf{\textit{Feature 2 ($\Psi_2$)}:} The subset of FBC bits that are flipped to logic 1.\\
\textbf{\textit{Feature 3 ($\Psi_3$)}:} The compression ratio ($r$) depends on the distribution of ones and zeros in a string (i.e., randomness). The compression ratio is defined as Eq. \ref{eqn:compRatio}.
\begin{equation} \label{eqn:compRatio}
r = \frac{S_{u}}{S_{c}}
\end{equation}

Where $S_{u}$ and $S_{c}$ are the sizes of the uncompressed and compressed data respectively. 

Our preliminary experimental results show that the compression ratio of $d_R$ varies from one manufacturer to another. We compress data using standard ZLIB library \cite{ZLIB:Deutsch} and then compared the data size with the original data. The ZLIB library is optimized for the minimal computational overhead while compressing the data. 

\textbf{\textit{Feature 4 ($\Psi_4$)}:} The whole block of $d_F$ is divided into a set of smaller blocks (each block is $64\times1$ of size and denoted by $B_w$). The standard deviation on the FBCs in these $B_w$s are considered as a feature. This feature captures the spatial locality of FBC along the dimension $w$. The higher value of standard deviation represents a greater spatial locality.\\
\textbf{\textit{Feature 5 ($\Psi_5$)}:} The block $d_F$ is divided into a smaller block, $B_b$ (of size $1\times8$) and then FBC is counted on each of the smaller blocks. Then we choose the standard deviation of those FBCs as the feature $\Psi_5$. The spatial locality along the dimension $b$ is captured with this feature $\Psi_5$.\\
\textbf{\textit{Feature 6 ($\Psi_6$)}:} Like $\Psi_4$, we calculate the standard deviation of FBCs on 64 blocks (of size $1024\times1$). This feature captures the fact that some cells of each 64-bit words are more error-prone than others.\\
\textbf{\textit{Feature 7 ($\Psi_7$)}:} Like $\Psi_5$, we calculate the standard deviation of FBCs on 1024 blocks (of size $1\times64$). This feature explores the fact that some bitlines are more error-prone than others.

All features except the $\Psi_2$ is extracted from all four data sets. The feature $\Psi_2$ is only extracted from the dataset 3 and dataset 4 (see Step 1). Choosing block-size for $\Psi_4$ through $\Psi_7$ is correlated to the DRAM organization. For example, we choose a block size of $1\times8$ ($B_b$) for $\Psi_5$ to capture the variations among the chips in a DRAM module. Our tested DRAM modules consist of four or eight chips, and each chip shares $1,024\times16$ or $1,024\times8$ blocks. We use block ($B_b$) height of 8 to extract average variations among the chips to ensure consistency among classes.

\textbf{Step 3: Machine-learning algorithms for detecting the DRAM origin.} 
After extracting the most suitable feature, we develop a machine-learning based technique to identify counterfeit DRAM modules. In our proposed technique, we use one-class classifier. Although one-class classifier is a more complex statistical problem, recent works demonstrated that it is more advantageous compared to other machine learning based techniques while detecting counterfeit ICs \cite{ML1:Huang, ML2:Sinanoglu, ML:Ahmadi, ML4:Huang}. On the other hand, in the traditional binary-class classifier, if the statistical diversity is enormous in the negative samples, the classifier might provide poor decision boundary due to the small negative train data \cite{SVDD:Chang, VDD:Tax}. In such a scenario, it will be very expensive or even impossible to collect data from the negative class covering the wide statistical diversity. This situation is particularly true for counterfeit IC detection as counterfeit ICs can be introduced from a wide variety of sources (see Sec. \ref{subsubsec:Motivations}). On the contrary, the one-class classifier \cite{SVDD:Chang, VDD:Tax, occSVM:Scholkoph, disOCC:Castillo, OCC:Taxonomy} is trained by only positive class samples. In our proposed method, we use Support Vector Data Description (SVDD) \cite{SVDD:Chang, VDD:Tax} to detect the outliers of a specific class. SVDD creates a spherical decision boundary in feature-space around the train dataset of a given class. For a given training data $x_{i} \in \mathcal{R}^n, i= 1, 2, 3, ..., l$, Tax et al. \cite{VDD:Tax} solved the following optimization problem given by Eq. \ref{eqn:svdd1}.

\begin{equation} \label{eqn:svdd1}
\begin{aligned}
& \min_{R,\bm{a},\xi} R^2 + \mathcal{C} \sum_{i=1}^{l} \xi_{i} \\
& \text{subject to, } \left \| \varphi(\bm{x})- \bm{a} \right \|^{2} \leq R^2 + \xi_{i}, i = 1, 2, 3, ..., l \\
&\hspace{1.7cm} \xi_{i} \geq 0, i = 1, 2, 3, ..., l
\end{aligned}
\end{equation}

Here, $\xi_{i}$ is a slack variable and $\varphi(\bm{x})$ is the mapping function from the lower dimension to a higher dimension. $R$ and $\bm{a}$ are the radius and center of the encircling boundary, and $\mathcal{C}$ is the regularization parameter. A smaller value of $\mathcal{C}$ causes more training samples to be treated as an outlier. A sample will be considered as an outlier if $\left \| \varphi(\bm{x})- \bm{a} \right \|^{2} > R^2$.
However, Eq. \ref{eqn:svdd1} can be efficiently solved by the Eq. \ref{eqn:svdd2}.

\begin{equation} \label{eqn:svdd2}
\begin{aligned}
& \max_{\alpha} \sum_{i} \alpha_{i}\mathcal{K}_{i,i} - \sum_{i,j} \alpha_{i}\alpha_{j}\mathcal{K}_{i,j} \\
& \text{where, } \sum_{i=1}^{l}\alpha_{i} = 1
\end{aligned}
\end{equation}

Here, $\mathcal{K}$ is the kernel function (i.e., $\mathcal{K}_{i,j} = \langle\varphi(x_{i})^{T}\cdot\varphi(x_{j})\rangle$). In our case, we have chosen the radial basis kernel function (Eq. \ref{eqn:radbas}) \cite{radbas:Scholkopf}. The radial basis function is useful when the data are not linearly separable.

\begin{equation} \label{eqn:radbas}
\mathcal{K}_{i,j} = exp(-\gamma\left \| x_{i} - x_{j} \right \|^2), \gamma > 0
\end{equation}

In Eq. \ref{eqn:radbas} $\gamma$ is a free parameter. A larger value of $\gamma$ enables the classifier to capture more complex attributes of the training data. On the other hand, the classifier model might suffer from overfitting problem if the value of $\gamma$ is too large. However, the $\mathcal{C}$ and $\gamma$ can be optimized more efficiently by introducing artificial outliers and applying k-fold cross-validation \cite{outlier:tax}. Moreover, the classifier accuracy can be increased by introducing some real negative examples during training \cite{VDD:Tax}.

\begin{figure}[ht!]
\vspace{-3mm}
\centering
\captionsetup{justification=centering, margin= 0cm}
\includegraphics[width=0.35 \textwidth]{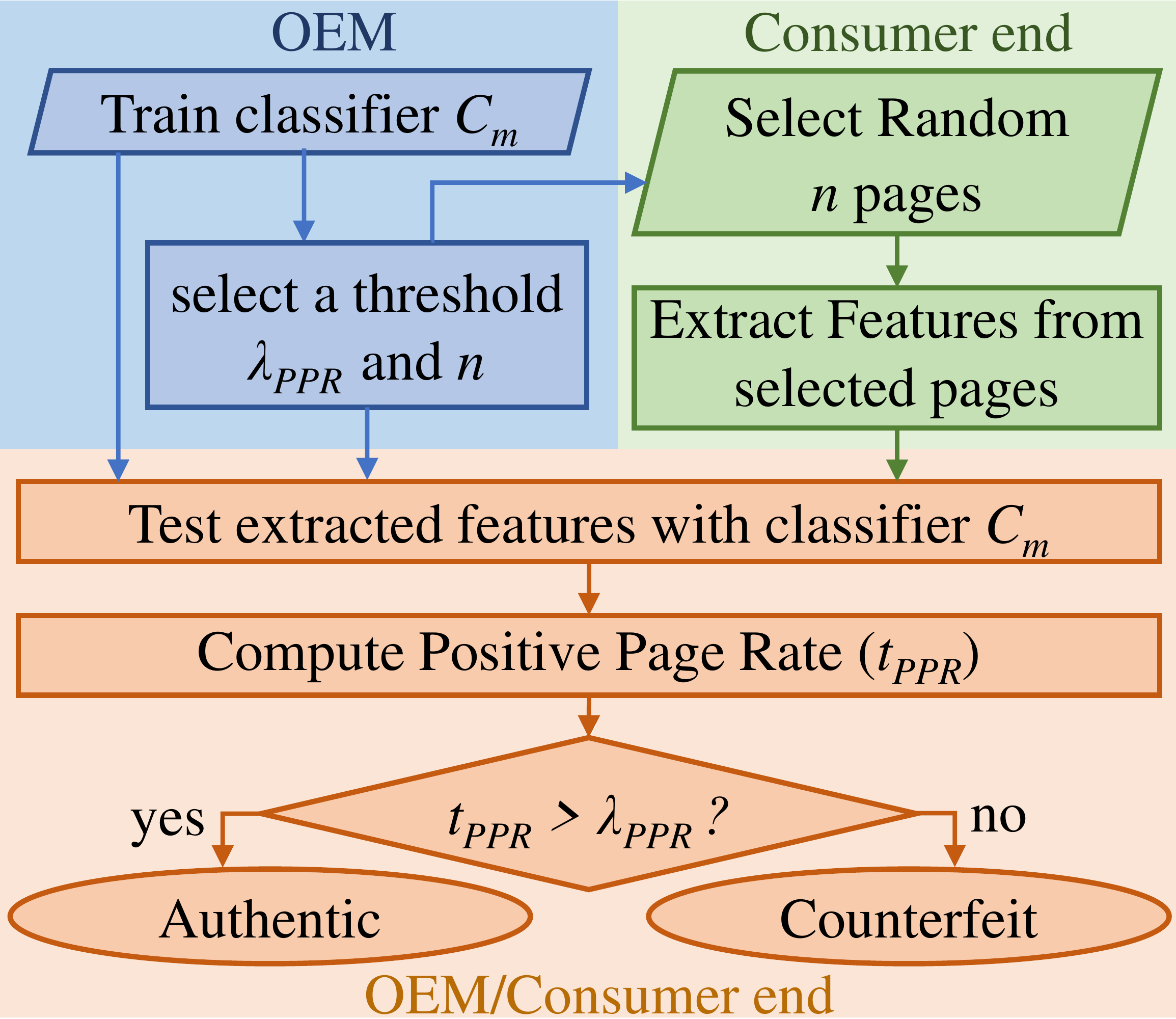} 
\caption{The proposed framework that is used to prove the authenticity of the origin of the DRAM manufacturer along with specification.}
\vspace{-3mm}
\label{fig:protocol}
\end{figure}

\textbf{Step 4: Constructing a framework to detect DRAM manufacturer.} Fig. \ref{fig:protocol} presents the proposed framework to identify the DRAM origin (i.e., the origin of the manufacturer and the specification of individual DRAM). In the proposed framework, we assume that the manufacturer or a trusted party provides the classifier model to the consumer and also defines a threshold for \textit{Positive Page Rate} ($PPR$). The \textit{positive page rate} ($PPR$) is defined as follows-

\begin{equation} \label{eqn:FPR}
\textit{PPR} = \frac{\textit{No. of pages that are classified as `positive'}}{\textit{No. of test pages from the memory module}}
\end{equation}

\begin{figure}[ht!]
\centering
\captionsetup{justification=centering, margin= 0cm}
\begin{subfigure}[t!]{0.38 \textwidth}
\includegraphics[width=\textwidth]{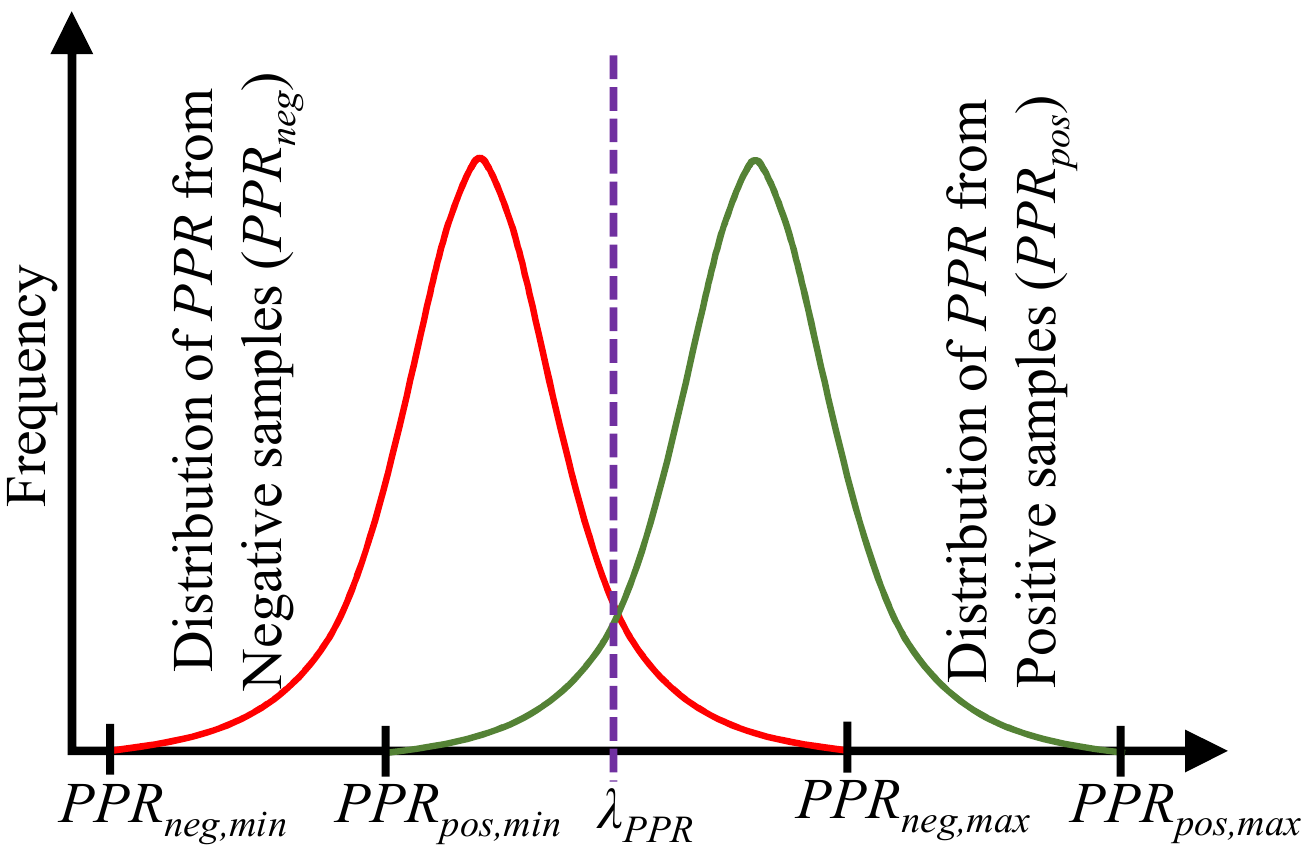} 
\caption{}
\label{fig:lamda_overlap}
\end{subfigure}
~
\begin{subfigure}[t!]{0.43 \textwidth}
\includegraphics[width=\textwidth]{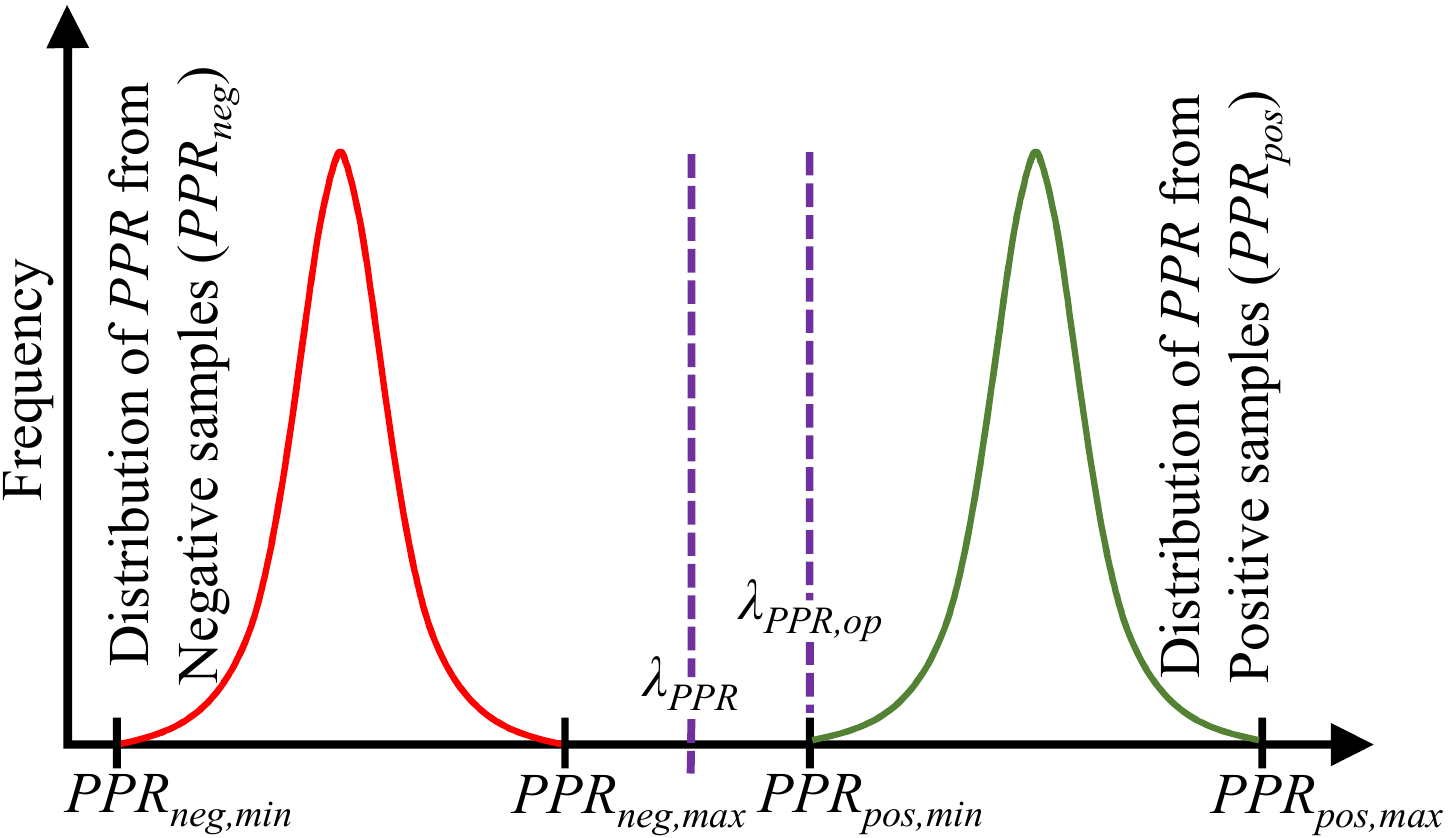}
\caption{}
\vspace{-2mm}
\label{fig:lamda_nonoverlap}
\end{subfigure}
\caption{Selecting $\lambda$ for case- (a) when the distribution of $PPR_{neg}$ and $PPR_{pos}$ are overlapped, (b) when the distribution of $PPR_{neg}$ and $PPR_{pos}$ are mutually exclusive. }
\vspace{-3mm}
\label{fig:lamda}
\end{figure}

In Fig. \ref{fig:protocol}, the steps, shown in the blue region, are performed by the OEM (Original Equipment Manufacturer), and the steps shown in the green region are performed at the consumer end. All other steps (covered by the orange region) can be processed in either consumers' system or manufacturers' system. Initially, the OEM or a trusted party trains a classifier based on all page data that are captured from one or multiple DRAM samples of the target class. The OEM or trusted party should also specify the number of memory pages that need to be tested from a memory module to prove its authenticity. Then, based on the sample statistics, the OEM should choose a threshold ($\lambda_{PPR}$) to decide whether a DRAM is manufactured by them or not. If the $PPR$ from the test module is higher than the threshold, then the memory module should be considered as authentic. The selection of the threshold value $\lambda_{PPR}$ and the number of test pages ($n$) depend on the quality of the classifier and the manufacturing process. Higher process variations might cause a large statistical diversity on the manufactured memory modules and may increase the chance of miss-classification. Besides, a larger process variation may lead to a higher statistical variation among the memory pages from the same DRAM module. In such a case, we might need more randomly sampled memory pages (i.e., a larger value of $n$) to capture all architectural and manufacturing process variations of a DRAM module. The choice of $\lambda_{PPR}$ mostly depends on the quality of classifier. Fig. \ref{fig:lamda_overlap} shows that the distribution of $PPR$ from positive samples and negative samples have an overlapping region. In such case, it is not possible to select a $\lambda_{PPR}$ that creates a clear boundary between the positive samples and the negative samples. On the other hand, if the distribution of the $PPR$ is mutually exclusive (Fig. \ref{fig:lamda_nonoverlap}), selecting a $\lambda_{PPR}$ within the interval [$PPR_{neg,max}$, $PPR_{pos,min}$] will separate positive and negative samples. Therefore, the ideal goal should be, maximizing the separation between $PPR_{neg,max}$ and $PPR_{pos,min}$ for a suitable value of $\lambda_{PPR}$ during classification. As it is difficult/impossible to collect the negative class data that covers the whole distribution (discussed in Step 3), the $\lambda_{PPR}$ should be defined with the highest possible value (i.e., $\lambda_{PPR,op} = PPR_{pos,min}$). In our proposed scheme, the OEM should train a classifier $C_m$, corresponding to a specific memory class and make the classifier parameter public. Then, the user should choose random $n$ test pages from the memory module that is under test. The general information given with the classifier $C_m$ should enable the user to extract features form those selected pages. Then, for each of those $n$ test pages, the OEM/user should test the extracted features using the classifier $C_m$. If the $PPR$ (calculated from Eq. \ref{eqn:FPR}) is higher than the threshold $\lambda_{PPR}$, the memory module should be marked as authentic. Otherwise, it should be identified as a counterfeit one.

\section{Results and Analysis} \label{sec:result}
In our experiment, we have collected data from 25 commercial off-the-shelf single rank DDR3 SODIMMs (small outline dual in-line memory module) from 3 major DRAM manufacturers (see Table 1) \cite{DRAMmarketShare}. We have tagged the memory class based on the part number, the Garber version (reference PCB layout version \cite{DesignFile:JEDEC}), and the SPD \cite{JEDEC:SPD} data (see Table 1). Among the first six classes, we found at least one mismatch in their SPD data (detailed of SPD data is not presented in this article). On the other hand, the last two classes only differed by their PCB layout version and SPD version. From each memory module, we have collected data from all memory banks (each DRAM module contains eight banks). The testing platform has been implemented using a Xilinx Virtex ML605 evaluation board with SoftMC \cite{SoftMC:Hasan}. Data have been written and fetched from the DRAM memory module with two 32-byte data bursts. For all memory modules, we have observed error patterns below $7.5ns$ of \textit{Activation} time (the recommended \textit{Activation} time is in between $10ns$ to $15ns$ \cite{DDR3:JEDEC}). In this work, we have conducted our experiment with a $5ns$ of \textit{Activation} time which should be achievable by most of the system. In order to quantify the robustness of our proposed technique, we have evaluated our proposed method in four different operating condition- i) nominal voltage ($1.5$v) and room temperature ($\ang{25}$C) (NVRT), ii) high voltage and room temperature (HVRT), iii) low voltage and room temperature (LVRT), and iv) nominal voltage and high temperature (NVHT). For HVRT and LVRT, we have changed the input voltage ($V_{DD}$) by $\pm{1}$\% as most of the memory controllers limit the voltage ripple within $\pm{1}$\% \cite{TI:MemPower}. For NVHT, we have changed the operating temperature by $+\ang{15}$C from the room temperature.

\begin{table}[ht!]
\setlength{\tabcolsep}{0.5em} 
\centering
\begin{tabular}{|c|c|c|c|c|c|}
\hline
\rule{0pt}{2ex} Manufacturer & \begin{tabular}[c]{@{}c@{}}Part\\ Name \footnotemark[2]\end{tabular} & \begin{tabular}[c]{@{}c@{}}Country\\ Origin\end{tabular} & Quantity & \begin{tabular}[c]{@{}c@{}}SPD-Garber\\ Version\end{tabular} & \begin{tabular}[c]{@{}c@{}}Class\\ tag\end{tabular} \\ \hline
\rule{0pt}{2ex} \multirow{3}{*}{Micron} & M1 & China & 2 & 10-C1 & 1 \\ \cline{2-6} 
\rule{0pt}{2ex} & \multirow{2}{*}{M2} & Singapore & 3 & \multirow{2}{*}{10-B1} & \multirow{2}{*}{2} \\ \cline{3-4}
\rule{0pt}{2ex} &  & China & 4 &  &  \\ \hline
\rule{0pt}{2ex} \multirow{4}{*}{Samsung} & S1 & China & 1 & 10-B1 & 3 \\ \cline{2-6} 
\rule{0pt}{2ex} & S2 & China & 1 & 11-B2 & 4 \\ \cline{2-6} 
\rule{0pt}{2ex} & \multirow{2}{*}{S3} & China & 4 & \multirow{2}{*}{11-B2} & \multirow{2}{*}{5} \\ \cline{3-4}
\rule{0pt}{2ex} &  & Philippines & 3 &  &  \\ \hline
\rule{0pt}{2ex} \multirow{3}{*}{SK Hynix} & \multirow{3}{*}{H1} & Korea & 5 & 10-B1 & 6 \\ \cline{3-6} 
\rule{0pt}{2ex} &  & China & 1 & \multirow{2}{*}{11-B2} & \multirow{2}{*}{7} \\ \cline{3-4}
\rule{0pt}{2ex} &  & Korea & 1 & & \\ \hline
\end{tabular}
\caption{Memory modules used in the data set.}
\vspace{-5mm}
\label{tab:dataInfo}
\end{table}
\footnotetext[2]{\textit{\textbf{M1:} MT4JSF12864HZ-1G4D1, \textbf{M2:}  MT8JSF12864HZ-1G4F1, \textbf{S1:} M471B2873EH1-CF8, \textbf{S2:} M471B2873GB0-CH9, \textbf{S3:} M471B5773DH0-CH9, \textbf{H1:} HMT325S6BFR8C-H9;\\\textbf{M1, M2, S2:} 1GB 1333MT/s, \textbf{S1:} 1GB 1066MT/s, \textbf{S3, H1:} 2GB 1333MT/s}}

\begin{figure*}[ht]
\centering
\captionsetup{justification=centering, margin= 0cm}
\begin{subfigure}[t!]{0.5\textwidth}
\includegraphics[width=\textwidth]{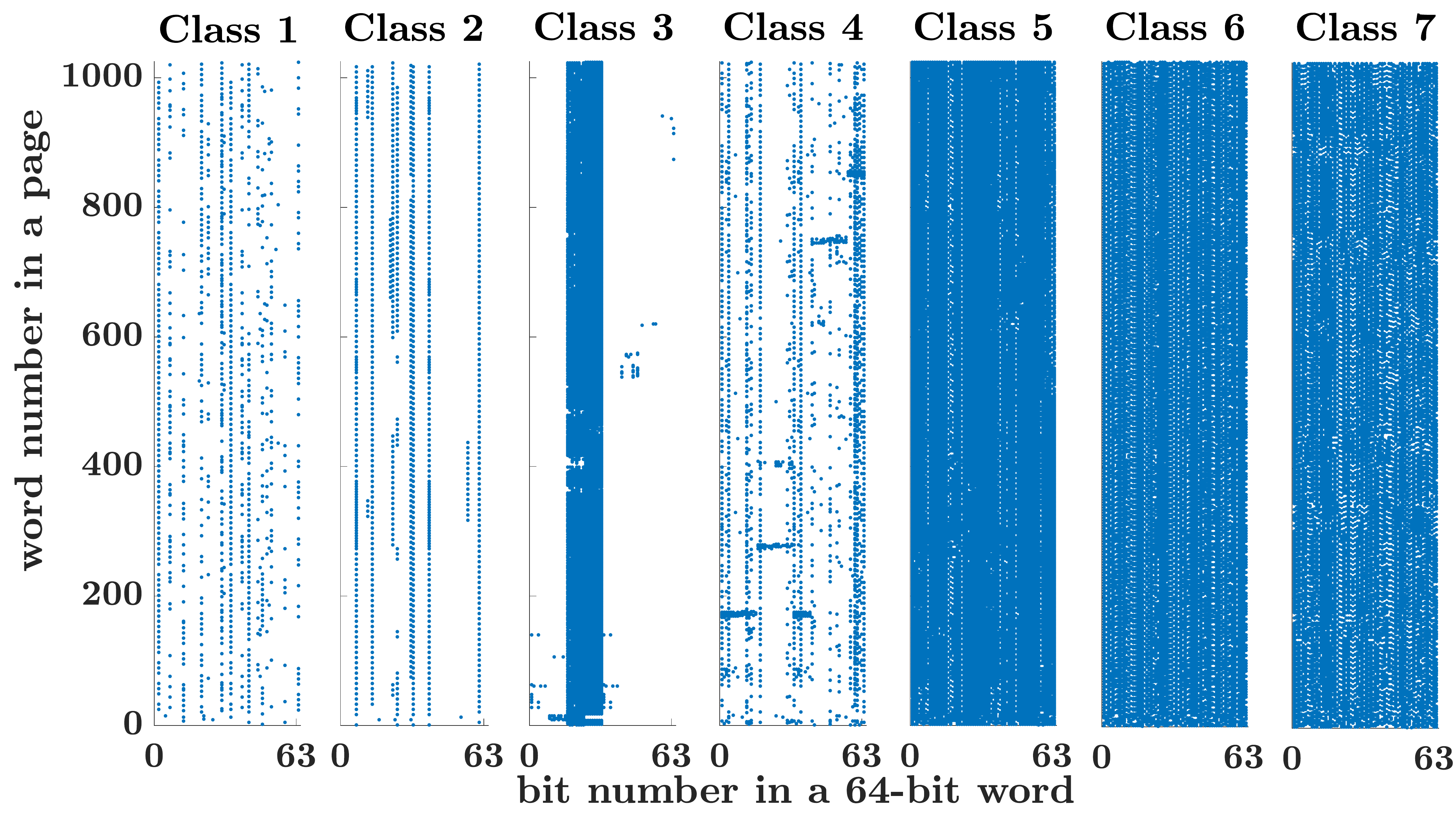} 
\caption{}
\vspace{-1mm}
\label{fig:pageEachClass}
\end{subfigure}
~
\begin{subfigure}[t!]{0.18\textwidth}
\includegraphics[width=\textwidth]{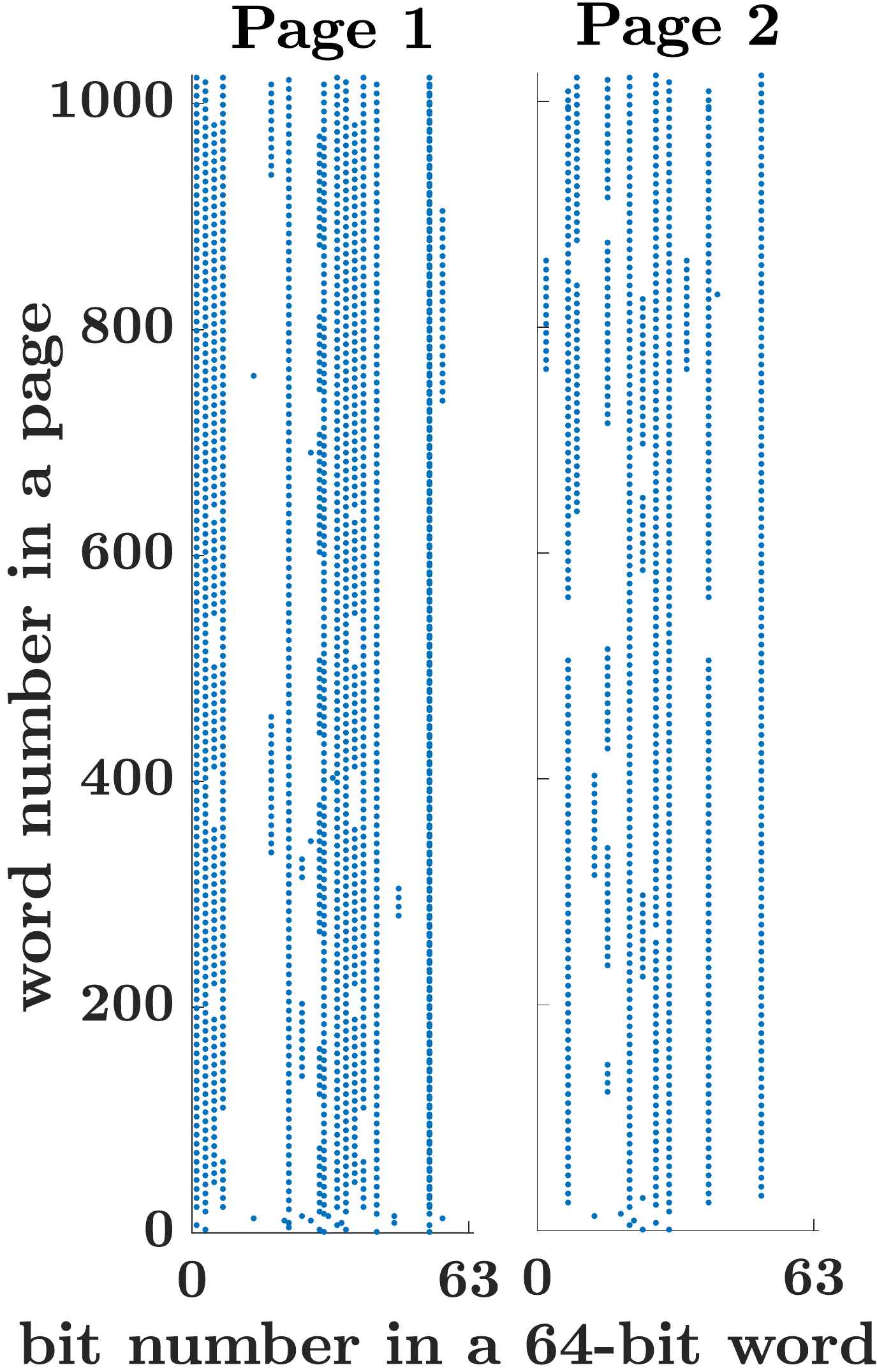}
\caption{}
\vspace{-1mm}
\label{fig:pageSameMemory}
\end{subfigure}
\caption{(a) Spatial locality of failed bits from a randomly chosen page from each class, \\(b) Spatial locality of randomly chosen two pages from same memory modules.}
\label{fig:patterenDistribution}
\end{figure*}

Fig. \ref{fig:patterenDistribution} presents the spatial locality of failed bits in a randomly chosen page form each memory class at NVRT operating condition. From the scatter plot, we observe that the error pattern is different for different classes. Note that, the PCB layout version only differs class 6 and class 7 and the subtle difference is difficult to understand from the figure (Fig. \ref{fig:pageEachClass}). In Fig. \ref{fig:pageSameMemory}, we have presented the spatial locality of failed bits on two random pages from the same memory module of the same class. Although there is some similarity in their texture, the pattern is not consistent. The features extracted from these samples (as discussed in Sec. \ref{sec:method}) are still capable of separating these classes.

\begin{figure*}[ht!]
\centering
\captionsetup{justification=centering, margin= 0.5cm}
\includegraphics[width=0.84\textwidth]{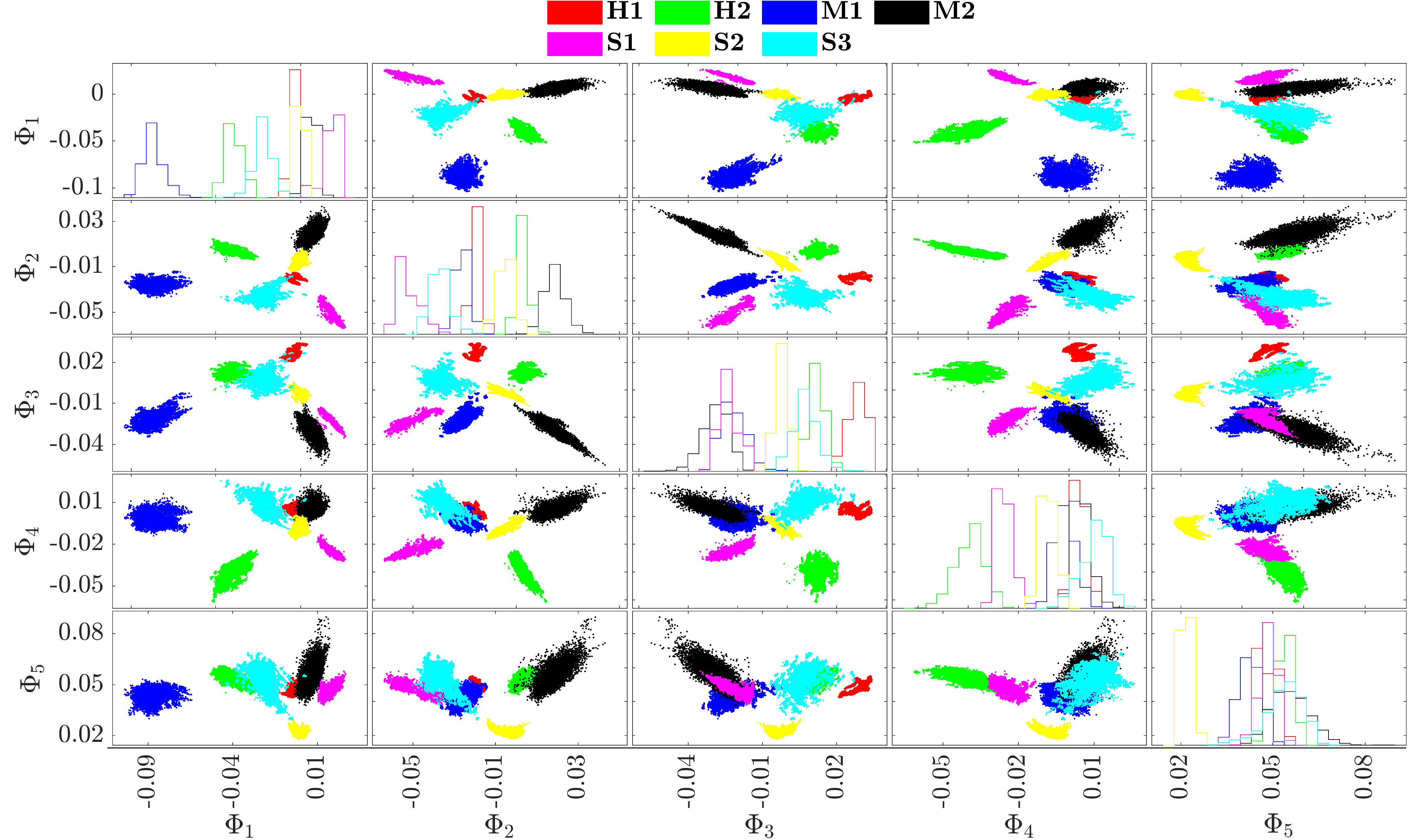} 
\caption{Visualizing data in feature-space.}
\vspace{-3mm}
\label{fig:visualizing}
\end{figure*}

From each memory page, we have extracted a total of 26 features as described in Sec. \ref{sec:method}. We have applied these 26 features directly to train and test the classifier. However, we have used the Linear Discriminant Analysis (LDA), a linear transformation \cite{LDA:Cunningham}, to provide the best visualization of the class separability. The LDA projects the data into a lower dimension feature-space by keeping the maximum separability among the classes. In the LDA, the lower dimension and the higher dimension features are linearly dependent. The data distribution in lower dimension feature-space is presented in Fig. \ref{fig:visualizing}. In this figure, we have only considered the most significant 5 dimensions ($\Phi_{1}, \Phi_{2}, \Phi_{3}, \Phi_{4},$ and $\Phi_{5}$) in the new feature-space that provides the maximum separability (explained variance). From the figure, we observe that each of the class forms a cluster in the feature space, which enables the separation of manufacturers.

To demonstrate our proposed method, we trained one-class SVDD classifier (as discussed in Sec. \ref{sec:method}) for each class where we assume that the manufacturer does not have any prior knowledge of the memory modules from other class (See Sec. \ref{sec:ProbDef}). The classifier was only trained at NVRT operating condition, and the same classifier was used to test the data from other operating conditions. Training one class classifier is a more complex statistical problem compared to the multi-class problem. We used LIBSVM library to implement the one class classifier \cite{libsvm:chang}. For each class, we selected only one module to train the classifier (using 26 features from all pages collected at NVRT condition) and then tested all the pages from the rest of the 24 modules with the trained classifier with all operating conditions.

To validate our algorithm, we have chosen all possible combinations of training and testing data sets. In Table \ref{tab:result}, we have presented the result from the one-class classifier. The third, fourth, and fifth columns of the table represent the mean, the standard deviation, and the minimum $PPR$ from the positive samples for each classifier ($PPR$ is calculated from each test module). The sixth, seventh, and eighth columns of the table represent the mean, the standard deviation, and the maximum $PPR$ from the negative samples. For the ideal case, the $PPR_{pos}$ and $PPR_{neg}$ should be 100\% and 0\% respectively. The standard deviation for both cases should be 0\%. A larger gap between the $PPR_{pos}$ and $PPR_{neg}$ provides us the flexibility of choosing appropriate values of $\lambda_{PPR}$ and $n$ for identifying the origin of the manufacturers along with specification with high confidence (Sec. \ref{sec:method}).
Our silicon results provide a satisfactory difference between the $PPR_{pos}$ and $PPR_{neg}$, which can be further improved by learning statistical model with more positive samples and/or introducing negative samples. Table \ref{tab:result} also presents that a small change in voltage and temperature has a very insignificant effect on classifier performance. This is expected because a small change in voltage or temperature has a negligible effect on \textit{Activation} time. \cite{DRAMLatencyPUF:Kim, DRAMlatency:Chang, reducedVoltage:chang, runtimePerformanceDRAM:Chandrasekar}.

\begin{table}[ht!]
\captionsetup{justification=centering, margin= 0cm}
\setlength\tabcolsep{7pt} 
\begin{tabular}{|c|c|c|c|c|c|c|c|}
\hline
\rule{0pt}{2ex} \rotatebox[origin=c]{90}{Class Tag} & \rotatebox[origin=c]{90}{\begin{tabular}[c]{@{}c@{}}Operating\\ Condition\end{tabular}} & \rotatebox[origin=c]{90}{$PPR_{pos, \mu}$} & \rotatebox[origin=c]{90}{$PPR_{pos, \sigma}$} & \rotatebox[origin=c]{90}{$PPR_{pos, min}$} & \rotatebox[origin=c]{90}{$PPR_{neg, \mu}$} & \rotatebox[origin=c]{90}{$PPR_{neg, \sigma}$} & \rotatebox[origin=c]{90}{$PPR_{neg, max}$} \\ \hline
\rule{0pt}{2ex} \multirow{4}{*}{1} & NVRT & 0.00 & 0.00 & 0.00 & 0.04 & 0.26 & 1.77 \\ \cline{2-8} 
\rule{0pt}{2ex} & HVRT & 0.00 & 0.00 & 0.00 & 0.03 & 0.20 & 1.37 \\ \cline{2-8} 
\rule{0pt}{2ex} & LVRT & 0.00 & 0.00 & 0.00 & 0.03 & 0.22 & 1.51 \\ \cline{2-8} 
\rule{0pt}{2ex} & NVHT & 0.00 & 0.00 & 0.00 & 0.05 & 0.31 & 2.08 \\ \hline
\rule{0pt}{2ex} \multirow{4}{*}{2} & NVRT & 99.07 & 1.16 & 95.40 & 0.00 & 0.02 & 0.17 \\ \cline{2-8} 
\rule{0pt}{2ex} & HVRT & 98.04 & 3.54 & 87.60 & 0.00 & 0.01 & 0.10 \\ \cline{2-8} 
\rule{0pt}{2ex} & LVRT & 99.09 & 0.98 & 96.49 & 0.00 & 0.02 & 0.19 \\ \cline{2-8} 
\rule{0pt}{2ex} & NVHT & 99.33 & 0.58 & 97.53 & 0.00 & 0.01 & 0.07 \\ \hline
\rule{0pt}{2ex} \multirow{4}{*}{3*} & NVRT & $-$ & $-$ & $-$ & 0.00 & 0.01 & 0.06 \\ \cline{2-8} 
\rule{0pt}{2ex} & HVRT & $-$ & $-$ & $-$ & 0.00 & 0.01 & 0.06 \\ \cline{2-8} 
\rule{0pt}{2ex} & LVRT & $-$ & $-$ & $-$ & 0.00 & 0.01 & 0.05 \\ \cline{2-8} 
\rule{0pt}{2ex} & NVHT & $-$ & $-$ & $-$ & 0.00 & 0.01 & 0.07 \\ \hline
\rule{0pt}{2ex} \multirow{4}{*}{4*} & NVRT & $-$ & $-$ & $-$ & 0.00 & 0.00 & 0.00 \\ \cline{2-8} 
\rule{0pt}{2ex} & HVRT & $-$ & $-$ & $-$ & 0.00 & 0.00 & 0.00 \\ \cline{2-8} 
\rule{0pt}{2ex} & LVRT & $-$ & $-$ & $-$ & 0.00 & 0.00 & 0.00 \\ \cline{2-8} 
\rule{0pt}{2ex} & NVHT & $-$ & $-$ & $-$ & 0.00 & 0.00 & 0.00 \\ \hline
\rule{0pt}{2ex} \multirow{4}{*}{5} & NVRT & 84.19 & 6.92 & 59.79 & 0.38 & 1.02 & 6.8 \\ \cline{2-8} 
\rule{0pt}{2ex} & HVRT & 84.74 & 7.16 & 59.31 & 0.41 & 1.10 & 7.19 \\ \cline{2-8} 
\rule{0pt}{2ex} & LVRT & 84.19 & 6.06 & 60.92 & 0.40 & 1.02 & 6.7 \\ \cline{2-8} 
\rule{0pt}{2ex} & NVHT & 82.56 & 7.63 & 57.00 & 0.38 & 1.03 & 6.96 \\ \hline
\rule{0pt}{2ex} \multirow{4}{*}{6} & NVRT & 90.29 & 9.31 & 69.58 & 1.58 & 4.16 & 25.91 \\ \cline{2-8} 
\rule{0pt}{2ex} & HVRT & 90.35 & 9.39 & 68.98 & 1.54 & 4.14 & 25.78 \\ \cline{2-8} 
\rule{0pt}{2ex} & LVRT & 81.98 & 9.49 & 69.74 & 1.69 & 4.52 & 28.65 \\ \cline{2-8} 
\rule{0pt}{2ex} & NVHT & 90.32 & 9.31 & 69.17 & 1.52 & 3.80 & 23.64 \\ \hline
\rule{0pt}{2ex} \multirow{4}{*}{7} & NVRT & 73.81 & 10.46 & 66.41 & 1.56 & 4.73 & 21.72 \\ \cline{2-8} 
\rule{0pt}{2ex} & HVRT & 74.22 & 9.36 & 68.11 & 1.61 & 4.91 & 22.98 \\ \cline{2-8} 
\rule{0pt}{2ex} & LVRT & 71.55 & 14.67 & 61.18 & 1.56 & 4.75 & 21.19 \\ \cline{2-8} 
\rule{0pt}{2ex} & NVHT & 76.01 & 6.53 & 71.39 & 1.53 & 4.72 & 22.15 \\ \hline
\end{tabular}
\vspace{1ex}
\rule{0pt}{2ex} {\raggedright *For class 3 and class 4, we have only one sample which is used to train the statistical model. There is no positive test sample left for these two classes.}
\caption{Results from the one-class classifier.}
\vspace{-6mm}
\label{tab:result}
\end{table}

\begin{figure*}[ht]
\centering
\captionsetup{justification=centering, margin= 0cm}
\begin{subfigure}[t]{0.245\textwidth}
\includegraphics[width=\textwidth]{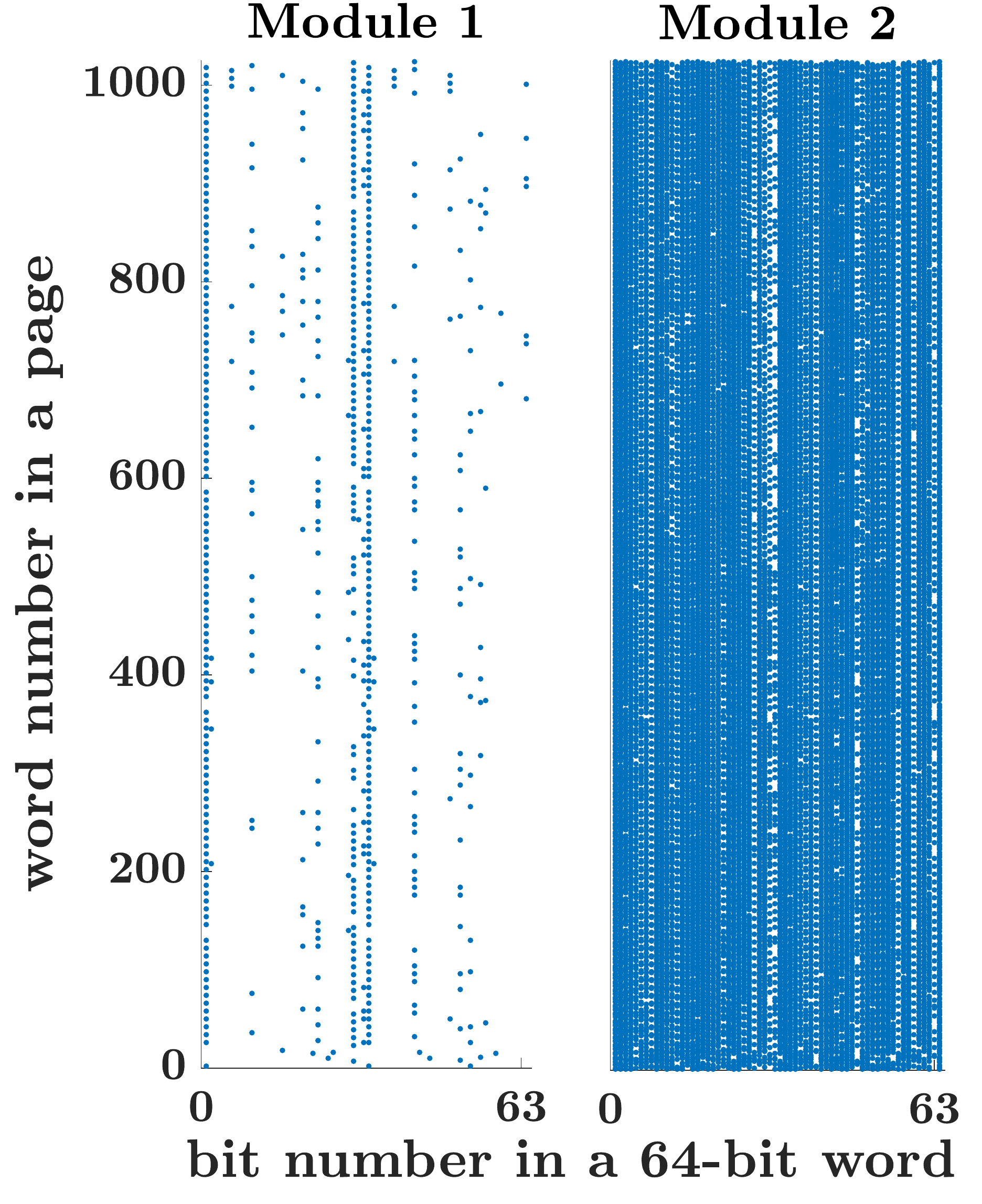} 
\caption{}
\label{fig:M1_justify_page2}
\end{subfigure}
~
\begin{subfigure}[t]{0.50\textwidth}
\includegraphics[width=\textwidth]{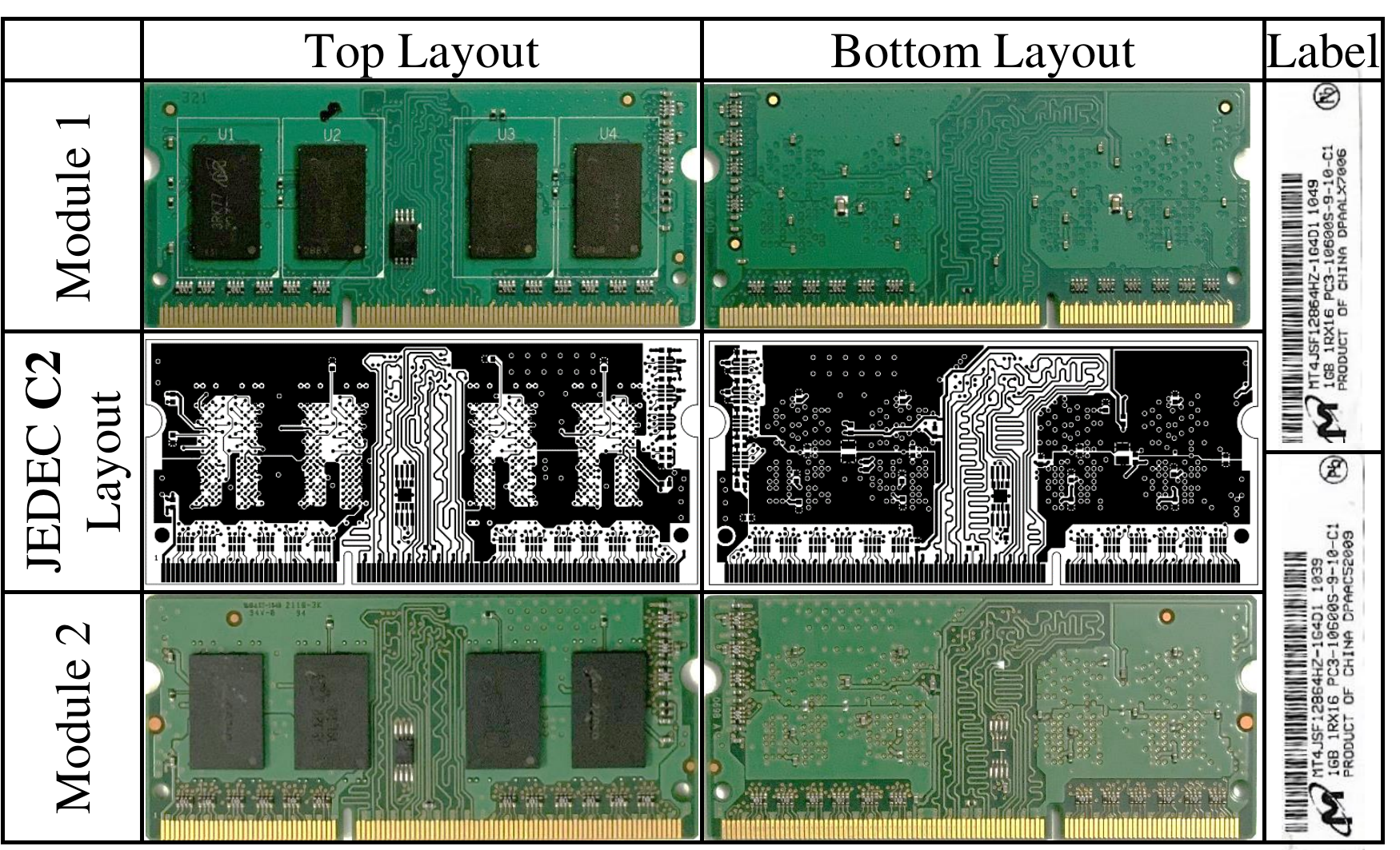}
\caption{}
\label{fig:M1_justify2}
\end{subfigure}
\caption{(a) Spatial locality of erroneous bits of 2 random pages of each sample from Class 1, \\ (b) Visual appearance of each sample from Class 1 (Module 2 is suspicious).}
\vspace{-3mm}
\label{fig:M1_justify}
\end{figure*}

\textbf{A suspicious DRAM module:} \label{subsec:suspiciousModule}
From the results shown in Table 3, for class 1, we observe that the $PPR_{pos}$ is 0\% (the ideal $PPR_{pos}$ is 100\%). Note that we have two samples available for this class: one is used for training, and another one is for testing. Therefore, we suspect that one of them is counterfeit. Fig. \ref{fig:M1_justify_page2} presents the spatial locality of failed bits from 2 random pages from those two samples. The results show that they have distinct FBC properties. Furthermore, the dissimilarities found in visual inspection (Fig. \ref{fig:M1_justify2}) suggests that one of them might be counterfeit (i.e., from a fake manufacturer). The layout difference between these two modules suggests that the reference layout version should be different for these two modules. However, the reference layout version is described as `C1' on both modules' label. From the SPD data, we have found that the reference raw card (i.e., layout) version is specified as `C' (which represents- `C0', `C1', `C2' etc.) for both modules. For further investigation, we have checked the layout provided by the JEDEC \cite{DesignFile:JEDEC} and found that the second module layout version is `C2' instead of `C1' (as shown in Fig. \ref{fig:M1_justify2}). Therefore, we conclude that the second module is either from the fake manufacturer or mislabeled (with layout version `C1').

\section{Limitations and Future Work} \label{sec:futureWork}

Our proposed method can be used to detect various counterfeit types. However, the proposed work cannot identify overproduction from the same foundry \cite{CSST_Rahman}. In the future, we will extend our technique for other volatile and non-volatile memory chips (e.g., flash memory, SRAM, etc. \cite{SRAM_rahman, flash_rahman}). Selecting a robust set of features to improve the accuracy might be another direction of our current research. In such a case, we might need to use more than one entropy source to capture a better variance among classes. In this article, we have only used data error obtained at the reduced activation latency.

Since our current work only exploits the learning ability of a one-class classifier, we will explore additional machine learning techniques, such as ensemble learning and classifier fusion, for improved generalization across a broader set of manufacturers in future work. We will also explore additional features and filtering, wrapper, and embedded feature analysis techniques to better understand the impact of each feature on identifying a DRAM manufacturer.   



\section{Conclusion} \label{sec:conclusion}
In this paper, we proposed a simple non-invasive and low-cost scheme for identifying the origin of a DRAM manufacturer and verifying individual DRAM's specification. The proposed method exploits the DRAM latency variations to capture the architectural, layout, and process variations. At first, we chose the most appropriate features from the DRAM signature, and then we used a one-class classifier to verify the memory class without knowing the information from other classes (i.e., other manufacturers).

\section*{Acknowledgment}
This work was supported by the National Science Foundation under Grant Number CNS-1850241. We would like to thank UAH for filing patent, UAH reference: UAH-P-18038.

\end{document}